\documentclass[11pt]{amsart}

\usepackage{epsfig}\usepackage{graphicx}
\usepackage{tikz}\usetikzlibrary{calc}

\usepackage{amsmath,amsthm,amssymb,amscd}

\usepackage{tikz}
\usetikzlibrary{arrows,shapes.geometric,positioning,decorations.markings}

\usepackage[normalem]{ulem}

\newcommand{\Id}{\operatorname{Id}}
\def\Mtwo{M_{\langle 2\rangle}}\def\Mthree{M_{\langle 3\rangle}}

 \def\nnn{\bold n}\def\mmm{\bold m}\def\lll{\bold l}
\def\Mn{M_{\langle \nnn \rangle}}

\def\Mthree{M_{\langle 3 \rangle}}
\def\ol{\overline}

\def\ttrace{\operatorname{trace}}

\def\FS{\mathfrak  S}
\def\BC{\mathbb C}\def\BF{\mathbb F}\def\cS{\mathcal S}
\def\BZ{\mathbb Z}

\def\ot{{\mathord{ \otimes } }}
\def\op{{\mathord{\,\oplus }\,}}
\def\ur{\underline {\bold R}}
\def\s{\sigma}

\def\t{\tau}\def\ur{\underline {\bold R}}\def\G{\Gamma}
\def\tinf{{\rm inf}}\DeclareMathOperator{\tlog}{log}

 \author{J.M. Landsberg}

\address{Department of Mathematics, Texas A\&M University, College Station, TX 77843-3368, USA}
\email{jml@math.tamu.edu}
 
\title[The complexity of matrix multiplication]{The complexity of matrix multiplication:\\ developments since 2014\\
Extended abstract of  2018  Oberwolfach Complexity meeting plenary lecture}

\thanks{Supported by NSF grants  DMS-1405348 and AF-1814254. This work was partially supported by the 
grant 346300 for IMPAN from the Simons Foundation and the matching 2015-2019 Polish MNiSW fund as well as
an  Simons Visiting Professor  grant supplied by the Simons Foundation
and by the Mathematisches Forschungsinstitut Oberwolfach. 
 }


\begin{document}
\maketitle
\thispagestyle{empty}

The complexity of all operations in linear algebra is governed by the complexity of  matrix multiplication.
In 1968 V. Strassen \cite{Strassen493}  discovered the way we usually multiply matrices is not
the most efficient one and initiated the central problem of determining 
the complexity of matrix multiplication. He defined  a fundamental
constant $\omega$, called the {\it exponent of matrix multiplication}, that governs its complexity.
For a tensor $T\in \BC^m\ot \BC^m\ot \BC^m$, let $\bold R(T)$ denote its tensor rank, the smallest $r$ such
that $T$ may be written as a sum of $r$ rank one tensors, and $\ur(T)$ its tensor border rank, the smallest
$r$ such that $T$ may be written as a limit of a sequence of rank $r$ tensors.
  Bini \cite{MR605920} proved the  border rank of the matrix multiplication
tensor $\ur(\Mn)$  asymptotically determines $\omega$. More precisely, considering $\ur(\Mn)$
as a function of $\nnn$, $\omega=\tinf_{\t}\{ \ur(\Mn)=O(\nnn^\t)\}$. 

This talk has two goals: (i) report on progress in the last four years
regarding upper and lower bounds for the complexity of matrix multiplication
and tensors in general, 
and (ii) to explain  the utility of algebraic geometry and representation theory for
matrix multiplication and complexity theory in general.

\subsection*{Lower bounds}
 Strassen-Lickteig (1983, 1985) \cite{Strassen505,MR86c:68040} showed $\ur(\Mn)\geq \frac {3\nnn^2}2+\frac\nnn 2 -1$.
Then,  after  $25$  years without progress,   Landsberg-Ottaviani (2013) \cite{MR3376667} showed $\ur(\Mn)\geq 2\nnn^2-\nnn$.
Around 2014 several authors \cite{2017arXiv171009502E,MR3781583,MR3611482}  independently proved that the existing lower
bound methods would not go much further. 
In \cite{MR3633766} the {\it border substitution method}  was developed, which led to the current best lower bound 
  $\ur(\Mn)\geq 2\nnn^2-\tlog_2(\nnn)-1$ \cite{MR3842382}. 

The geometric approach to lower bounds is as follows:
let $  \s_r:=\{ T\in \BC^m\ot \BC^m\ot \BC^m \mid \ur(T)\leq r\}$, the set of tensors in $\BC^m\ot \BC^m\ot \BC^m$ of border rank at most $r$.
This set is an {\it algebraic variety}, i.e., it  is the zero set of a collection of (homogeneous) polynomials. 
Na\"\i vely expressed, to prove   $\ur(\Mn)>r$, or to prove lower border rank bounds for any tensor, one simply looks for a polynomial in the ideal
of $\s_r$ (that is a polynomial $P$ such that $P(T)=0$ for all $T\in \s_r$) such that $P(\Mn)\neq 0$ (here $m=\nnn^2$).
But   how can one find such polynomials? This is where {\it representation theory} comes in. The variety
$\s_r$ is {\it invariant} under changes of bases in the three spaces. That is, write $\BC^m\ot \BC^m\ot \BC^m=A\ot B\ot C$,
and let  $GL(A)$ etc.. denote the invertible $m\times m$ matrices. There is a natural action of 
$G:=GL(A)\times GL(B)\times GL(C)$ on $A\ot B\ot C$:  on rank one tensors $(g_A,g_B,g_C)\cdot (a\ot b\ot c):=(g_Aa)\ot (g_Bb)\ot (g_Cc)$, and
the  action on $A\ot B\ot C$ is defined by extending this action linearly. Then for all $g\in G$ and $x\in \s_r$, one has $g\cdot x\in \s_r$.
Whenever  a variety is invariant under the action of a group, its   ideal  
is invariant under the group as well via the induced
action on polynomials. One can then attempt to use representation theory to decompose the space of all polynomials
and  systematically check which irreducible modules are in the ideal. This works well in small dimensions, e.g.,
to show $\ur(\Mtwo)=7$ \cite{MR3171099}, but in general one must use additional  methods. A classical approach
is to try to embed $A\ot B\ot C$ into a space of matrices, and then take minors, which (in a slightly different context) dates
back at least to Sylvester. The advance here was to look for $G$-{\it equivariant} ($G$-{\it homomorphic}) embeddings.
This idea led to the 2013 advance, but the   limits described in \cite{2017arXiv171009502E,MR3781583,MR3611482} exactly apply to such embeddings, so to advance
further one must find new techniques.

At this point I should mention the general {\it hay in a haystack} problem of finding explicit
sequences of tensors $T_m\in \BC^m\ot \BC^m\ot \BC^m$ of high rank and
border rank. The maximum border rank is $\lceil \frac{m^3}{3m-2}\rceil$ for $m>3$, and
the maximum possible rank is {\it not known}. The current state of the art are
explicit tensors with $\bold R(T_m)\geq 3m-o(m)$ \cite{MR3025382,MR3162411} and others with $\ur(T_m)\geq 2m-2$ \cite{MR3320240}.
The border substitution method promises to at least improve the state of the art on these problems.

A last remark on lower bounds: in the past two months there has been a very exciting breakthrough due to
Buczynska-Buczynski (personal communication), that avoids the above-mentioned barriers in the polynomial
situation. The method is a combination of the
classical apolarity method  with the border substitution method.
Buczynski and I are currently working to extend these methods to the tensor situation.

\subsection*{Upper bounds}
The {\it Kronecker product} of $T\in A\ot B\ot C$ and $T'\in A'\ot B'\ot
C'$, denoted  $T\boxtimes
T'$, is the tensor $T \ot T' \in (A\ot A')\ot (B\ot B')\ot (C\ot C')$
regarded as $3$-way tensor. The Kronecker powers of $T$, $T^{\boxtimes
N}\in (A^{\ot N})\ot (B^{\ot N})\ot (C^{\ot N})$ are defined similarly. Also let $T\op T'\in (A\op A')\ot (B\op B')\ot (C\op C')$ 
denote the direct sum. Let $M_{\langle \lll,\mmm,\nnn\rangle}$ denote the rectangular matrix multiplication tensor. 
The matrix multiplication tensor has the remarkable property that $M_{\langle \lll,\mmm,\nnn\rangle} \boxtimes
M_{\langle \lll',\mmm',\nnn'\rangle}=M_{\langle \lll\lll',\mmm\mmm',\nnn\nnn'\rangle}$ which is the key to Strassen's {\it laser method}.

Following work of Strassen, Bini  \cite{MR605920} showed that for all $\lll,\mmm,\nnn$,  setting $q=(\lll\mmm\nnn)^{\frac 13}$, that 
$$
\omega\leq \frac{\tlog(\ur(M_{\langle \lll,\mmm,\nnn\rangle}))}{\tlog(q)}.
$$
This was generalized by Sch\"onhage \cite{MR623057}. A special case is as follows: say that   for $1\leq i\leq s$, $\lll_i\mmm_i\nnn_i= q^3$,
then 
$$
\omega\leq \frac{\tlog(\frac 1s  \ur(\bigoplus_{i=1}^s M_{\langle \lll_i,\mmm_i,\nnn_i\rangle}))}{\tlog(q)}.
$$ 
Sch\"onhage  also showed that border rank can be strictly sub-additive, so the result is nontrivial. This immediately
implies that if $T$ is a tensor such that $\bigoplus_{i=1}^s M_{\langle \lll_i,\mmm_i,\nnn_i\rangle}\in \ol{G\cdot T}$,
i.e., $T$ {\it degenerate}s to $\bigoplus_{i=1}^s M_{\langle \lll_i,\mmm_i,\nnn_i\rangle}$, 
then
$$
\omega\leq \frac{\tlog(\frac 1s  \ur(T))}{\tlog(q)}.
$$ 
This  can be useful if the border rank of $T$ is easier to estimate than that of the direct sum of matrix multiplication
tensors. One should think of $\ur(T)$ as the {\it cost} of $T$ and $s$ and $\tlog(q)$ as determining
the {\it value} of $T$. One gets a good upper bound if cost is low and value is high. Strassen \cite{MR882307}  then showed that the same result holds if the
matrix multiplication tensors are nearly disjoint (i.e., nearly direct sums)  by taking Kronecker powers of $T$ and degenerating the powers to disjoint matrix multiplication
tensors.
 This method was used by  Coppersmith and Winograd \cite{copwin135} with the now named
\lq\lq little Coppersmith-Winograd tensor\rq\rq : 
$$T_{cw,q}:=\sum_{j=1}^q a_0\ot b_j\ot c_j + a_j\ot b_0\ot c_j+ a_j\ot b_j\ot c_0, 
$$
to show
$$
\omega \leq \frac{\tlog(\frac 4{27}  \ur(T_{cw,q})^3)}{\tlog(q)}.
$$
Since $\ur(T_{cw,q})=q+2$, this  implies $\omega <2.41$ when $q=8$.
They improved this to $\omega\leq 2.3755$ using a slightly more complicated tensor (the Kronecker square of the
big Coppersmith-Winograd tensor $T_{CW,q}$) which held the world record
until 2012-3, when it was lowered to $\omega\leq 2.373$ \cite{stothers,williams,LeGall:2014:PTF:2608628.2608664} using higher
Kronecker powers of $T_{CW,q}$.
Then in 2014, Ambainus, Filmus and LeGall \cite{MR3388238} proved that coordinate restrictions of 
Kronecker powers of  the big Coppersmith-Winograd tensor could never be used
to prove $\omega<2.3$. This was generalized in \cite{DBLP:conf/innovations/AlmanW18,2018arXiv181008671A} to a larger 
class of tensors and degenerations, albeit with weaker bounds on the limitations. Regarding Kronecker powers, one also has
\begin{equation} \label{cwk}
\omega \leq \frac{\tlog(\frac 4{27}  \ur(T_{cw,q}^{\boxtimes k})^{\frac 3k}) }{\tlog(q)},
\end{equation}
for any $k$.

I point out that all the above has little to do with practical matrix multiplication, the only better practical decomposition
to arise since Strassen's 1968 work is due to V. Pan \cite{MR765701}.

Given the   barriers from \cite{MR3388238,DBLP:conf/innovations/AlmanW18,2018arXiv181008671A}, it makes sense to ask what geometry can do for upper bounds.

\smallskip

{\it First idea}: study known rank decompositions of $\Mn$ to obtain new ones.
For a tensor $T$, let 
 $G_T:=\{ g\in G\mid g\cdot T=T\}$, denote the {\it symmetry group}
of $T$. For example $G_{\Mn}$ is the image of $GL_\nnn^{\times 3}$ in $GL_{\nnn^2}^{\times 3}$.
In \cite{CILO,BILR} we studied decompositions and noticed that many of the decompositions had large {\it symmetry groups}, where
if $\cS_T$ is a rank decomposition of a tensor $T$, if one applies an element of $G_T$
to the decomposition, it takes
it to another rank decomposition of $T$, which sometimes is the same as the original.
Let $\G_{\cS_T}:=\{ g\in G_T\mid g\cS_T=\cS_T\}$, denote the symmetry group of the decomposition.
Then the decomposition may be expressed in terms of the orbit structure. For example,  Strassen's 1968 rank $7$ decomposition of $\Mtwo$ may be written
$$
\Mtwo= \Id^{\ot 3} + \G\cdot\left[ \begin{pmatrix} 0&0\\ 1&0\end{pmatrix} \ot \begin{pmatrix} 0 & 0\\ 1&1\end{pmatrix}
\ot \begin{pmatrix} 0&1\\ 0 &-1\end{pmatrix} \right]
$$
where $\G\simeq \FS_3\rtimes \BZ_2$, and $\Gamma \cdot$ denotes the sum of the terms
in the $\G$-orbit. Here   the orbit consists of six terms. The identity is acted on trivially by $\G$, so the
decomposition is a union of two orbits.
This is work in progress.

\smallskip

{\it Second idea}: expand the playing field. Given any tensor, one can symmetrize it to get a cubic polynomial.
Let $s\Mn$ denote the symmetrized matrix multiplication tensor, the polynomial $X\mapsto \ttrace(X^3)$.
In \cite{MR3829726} we showed that the Waring   rank of $s\Mn$  also governs
the exponent of matrix multiplication, where the Waring rank of a cubic polynomial is the smallest $r$ such
that the polynomial may be written as a sum of $r$ cubes.

\smallskip

{\it Third idea}: combine the first two. A Conner \cite{2017arXiv171105796C} found a remarkable Waring rank $18$ decomposition of $s\Mthree$, with
symmetry group that of the Hasse diagram, namely $(\BZ_3^{\times 2}\rtimes SL_2(\BF_3))\rtimes \BZ_2$.   He
also found a Waring rank $40$ decomposition of $sM_{\langle 4\rangle}$ with symmetry group that
of the cube. For comparison, the best known rank decompositions of $\Mthree,M_{\langle 4\rangle}$
respectively are of ranks $23$ and $49$. This launched his program to find explicit sequences of finite groups $\G_\nnn\subset G_{s\Mn}$ such that
the space of $\G_\nnn$-invariants in the space of cubic polynomials in $\nnn^2$ variables only contains polynomials of low Waring rank,
translating the study of upper bounds on $\omega$  to a study of properties of sequences of finite groups, in the spirit of (but very
different from) the Cohn-Umans program \cite{CU}.

\smallskip

{\it Fourth idea}: the Ambainus-Filmus-LeGall challenge: find new tensors useful for the laser method. 
Michalek and I \cite{MR3682743} had the idea to isolate {\it geometric} properties of the Coppersmith-Winograd
tensors  and  to find other tensors with similar geometric properties,  in the hope that they might also be useful for the laser method.
We succeeded in isolating many interesting geometric properties. Unfortunately, we then proved that the Coppersmith-Winograd
tensors were the {\it unique} tensors with such properties.

\smallskip

{\it Fourth idea, second try}: In \cite{CGLV} we examine  the symmetry groups of the Coppersmith-Winograd tensors. We found that the big Coppersmith-Winograd
tensor has the largest dimensional symmetry group among {\it $1$-generic tensors} in odd dimensions
 ($1$-genericity is a natural condition for implementing the laser method),
but that in even dimensions, there is an even better tensor, which we call the {\it skew big Coppersmith-Winograd tensor}. We also found
other tensors with large symmetry groups. Unfortunately, none of the new tensors with maximal or near maximal symmetry groups
are better for the laser method than $T_{CW,q}$.

\smallskip

{\it Fifth idea}: Go back to the inequality \eqref{cwk} and upper bound Kronecker powers of $T_{cw,q}$ (this was posed as an open question
for the square as early as \cite{blaserbook}), which could even (when $q=2$) potentially show $\omega=2$. Unfortunately,  we show  in \cite{CGLV2} that
$15\leq \ur(T_{cw,2}^{\boxtimes 2})\leq 16$, and we expect $16$, and for $q>2$, that  $\ur(T_{cw,q}^{\boxtimes 2})=(q+2)^2$.

\smallskip

{\it Sixth idea}: Combine the last two ideas. The skew cousin of the little Coppersmith-Winograd tensor also satisfies   \eqref{cwk}.
It has the same value, but unfortunately, it has higher cost. For example,   $\ur(T_{skew-cw,2})=5>4=\ur(T_{cw,2})$.
However, we show  it satisfies $\ur(T_{skew-cw,2}^{\boxtimes 2})=17\ll 25=\ur(T_{skew-cw,2})^2$. This is one of the few  explicit tensors
known 
to have strictly submultiplicative border rank under Kronecker square (the first known being $\Mtwo$), and if this drop continues, it would
be very good indeed for the laser method.

\bibliographystyle{amsplain}

\bibliography{Lmatrix}

\end{document}